\documentclass[final]{elsarticle}
\usepackage{changes}
\usepackage{lineno,hyperref}
\usepackage{amsmath}
\usepackage{amsfonts}
\usepackage{gensymb}
\usepackage{hyperref}
\usepackage{subcaption}
\usepackage{xcolor}
\usepackage{chngcntr}
\usepackage{pdfpages}
\counterwithout{figure}{subsection}
\usepackage{siunitx}
\usepackage{multirow}
\usepackage{changes}
\modulolinenumbers[1]

\begin{document}

\begin{frontmatter}

\title{Generalized Method for the Optimization of Pulse Shape Discrimination Parameters}

\author[mymainaddress]{J. Zhou}
\author[mysecondaryaddress]{A. Abdulaziz}
\author[mysecondaryaddress]{Y. Altmann}
\author[mymainaddress]{A. Di Fulvio
\corref{mycorrespondingauthor}}
\cortext[mycorrespondingauthor]{Corresponding author.}
\ead{difulvio@illinois.edu}

\address[mymainaddress]{Department of Nuclear, Plasma, and Radiological Engineering, \\University of Illinois at Urbana-Champaign, Urbana, IL, USA}
\address[mysecondaryaddress]{School of Engineering and Physical Sciences, Heriot-Watt University, Edinburgh, UK}

\begin{abstract}
Organic scintillators exhibit fast timing, high detection efficiency for fast neutrons and pulse shape discrimination (PSD) capability. PSD is essential in mixed radiation fields, where different types of radiation need to be detected and discriminated. In neutron measurements for nuclear security and non proliferation effective PSD is crucial, because a weak neutron signature needs to be detected in the presence of a strong gamma-ray background. The most commonly used deterministic PSD technique is charge integration (CI). This method requires the optimization of specific parameters to obtain the best gamma-neutron separation. These parameters depend on the scintillating material and light readout device and typically require a lengthy optimization process and a calibration reference measurement with a mixed source. In this paper, we propose a new method based on the scintillation fluorescence physics that enables to find the optimum PSD integration gates using only a gamma-ray emitter. 
We demonstrate our method using three organic scintillation detectors: deuterated trans-stilbene, small-molecule organic glass, and EJ-309. In all the investigated cases, our method allowed finding the optimum PSD CI parameters without the need of iterative optimization.
\end{abstract}


\end{frontmatter}

%\linenumbers

\section{Introduction}

Pulse-shape-discrimination (PSD) capable organic scintillators are the detectors of choice when it is necessary to detect and discriminate different radiation types, e.g., gamma rays and neutrons, with fast timing and high efficiency. 
Therefore, organic scintillators are used for a wide range of applications, from nuclear security to diagnostic radiology and nuclear physics \cite{DIFULVIO201792,Shin2017246,Kimt200231,Kim199929}. The dependence of the fluorescence time constants on the particle linear energy transfer (LET) enables PSD \cite{Knoll}. In practice, PSD is possible because the shape of the detected pulses changes with the LET of the particle depositing its energy in the detector. 
The traditional and most commonly used method to find a parameter that depends on the pulse shape and hence enables PSD is based on charge integration (CI)\cite{Knoll,Ranucci1995389}. The CI-based PSD parameter is the tail-to-total ratio (TTR), which is the ratio between the area under the terminal portion of the pulse, i.e., the pulse tail, and the whole pulse area. TTR is calculated for each pulse and ranges between zero and one. A relatively high TTR corresponds to pulses with an increased delayed fluorescence emission with respect to the prompt fluorescence emission. Despite the increasing number of alternative PSD approaches, e.g., based on zero-crossing \cite{NAKHOSTIN20121,NAKHOSTIN2010498,SPERR197455,Roush1964112, Kuchnir4324923}, time-over-threshold \cite{Roy2022} and machine-learning \cite{Fu2018410, Abdulaziz20223538}, the CI remains the most frequently-used method for PSD.
While being simple to implement, CI requires a lengthy, source- and material-specific optimization of the pulse integration gate parameters, which is typically performed by evaluating the PSD figure-of-merit \cite{Langeveld20171801} for many iterations of such parameters. 

In this work, we present a generalized method to optimize the choice of the CI gate parameters. The proposed method is based on an exponential model to fit a template gamma-ray pulse. The model includes only the prompt fluorescence component, without accounting for the delayed fluorescence. The pulse time stamp at which the model and the measured template differ the most reveals the onset of the delayed component, which can be used as tail start time in CI-based PSD. 
This method is based on the intrinsic scintillation decay times, hence avoids the cumbersome gate optimization process needed in CI PSD. Moreover, it does not require any neutron source to find the optimized charge integration parameter. We validated the method using three organic scintillators: deuterated trans-stilbene (stilbene-d$_{12}$) \cite{ZHOU2022166287}, EJ-309, and small-molecule organic glass \cite{CARLSON2016152}, hereafter referred to as organic glass detector. The model-determined tail start time yielded the best PSD results for all three organic scintillation detectors.\par

\section{Methods}

We present a model-based method to find the integration time gates for CI PSD without the need for any iterative optimization process. \replaced{}{We validated} This method \added{was validated} using three organic scintillation detectors, namely, stilbene-d$_{12}$, EJ-309, and organic glass. 

\subsection{Workflow of the model-based charge integration PSD}

\begin{figure}[htbp!]
    \centering
    \includegraphics[width=0.7\textwidth]{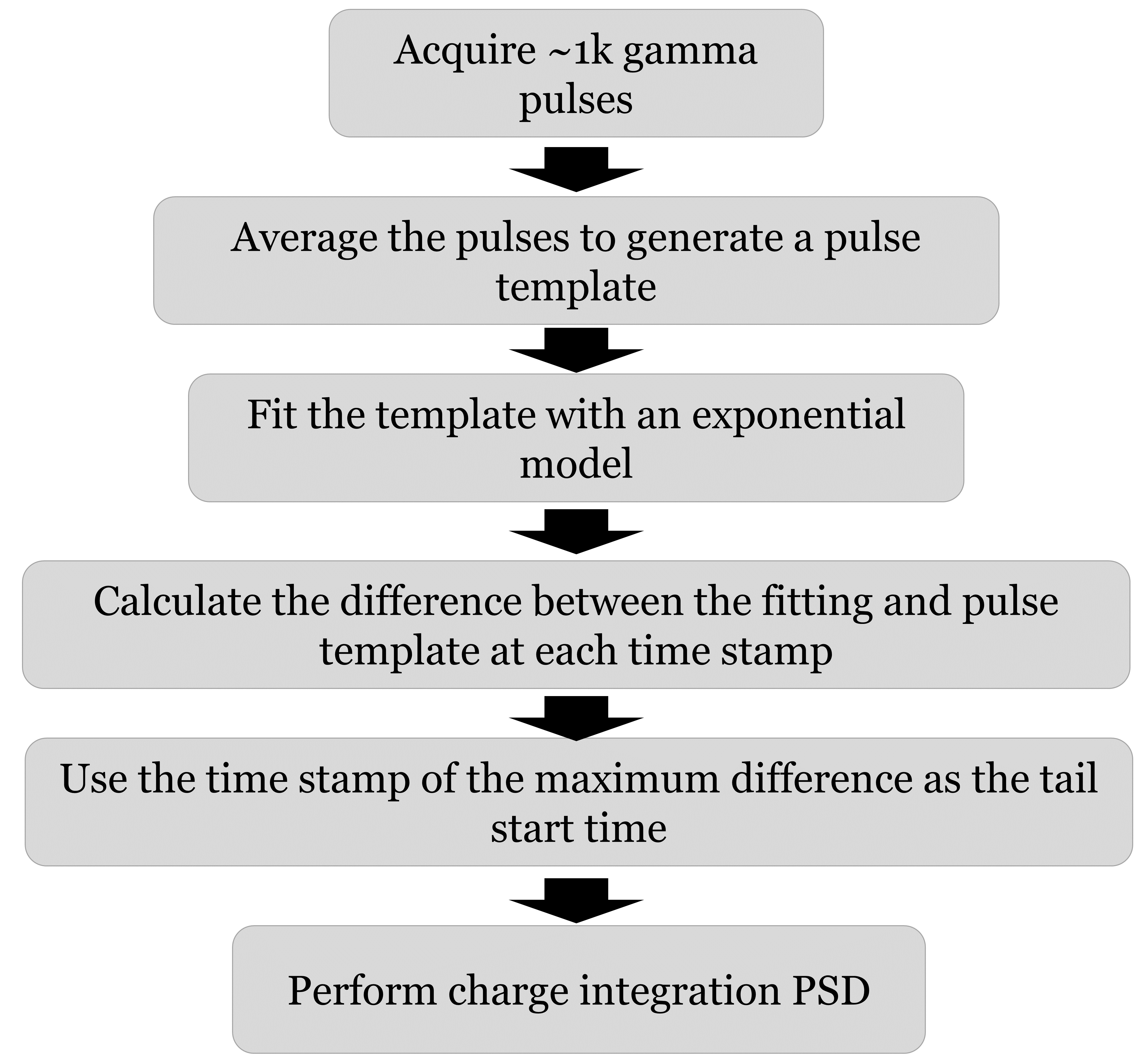}
    \caption{Workflow of the model-based charge integration method for neutron/gamma discrimination}
    \label{f:m:cfd_workflow}
\end{figure}

Figure \ref{f:m:cfd_workflow} shows the workflow of the proposed method to find the integration parameters for a PSD-capable scintillator. \replaced{}{We} 
First, \replaced{}{acquired} approximately one thousand gamma-ray pulses 
\added{were acquired} and averaged \replaced{}{them} to generate a pulse template. The details of the average process are described in the next section. Then, \replaced{}{we used} an exponential model \added{was used} to fit the template. After the fitting process, \replaced{}{we calculated} the pulse height differences between the original pulse template and the fitting result \added{were calculated}. The time stamp corresponding to the maximum difference between the two is the optimum tail start time for charge integration PSD. The details of this procedure are presented in the following sections.  

%\subsection{Constant Fraction Discrimination}

%Figure \ref{f:m:cfd} shows the CFD process that converts unipolar pulses to bipolar pulses and generates zero-crossing points. It is a widely-used timing method that provides a time marker for the signal and eliminates amplitude-dependent time walk for signals having consistent rise times \cite{GEDCKE1967377}. CFD entails the sum of an attenuated version of the pulse with the same pulse inverted and delayed. Therefore, it is based on two parameters: the attenuation fraction \textit{f} and the delay time \textit{$\Delta$}. We performed CFD on detector pulses and used the zero-crossing point as the tail start time to perform charge integration PSD.

%\begin{figure}[htbp!]
%    \centering
%    \includegraphics[width=0.85\textwidth]{figure/cfd_schematic.jpg}
%    \caption{ Constant fraction discrimination of signal processing}
%    \label{f:m:cfd}
%\end{figure}

%The CFD delay time directly affects the location of the zero-crossing point, i.e., the tail start time of charge integration PSD in our case. The procedure to calculate the optimum delay time \textit{$\Delta$} is outlined in the next section. 

\subsection{Derivation of CI integration time gate from the pulse fit}
\label{s:m:exponential_fit}

\replaced{}{We used the} Measured gamma-ray pulses and an exponential \added{pulse} model \added{were used} to obtain the optimal tail start time for CI PSD method. We first acquired approximately one thousand gamma-ray pulses and averaged them to a pulse template, normalized to its peak value. Then, we fit the template with a bi-exponential pulse model that is widely used to describe the fluorescence signal produced by organic scintillators \cite{MARRONE2002299}. The model is shown in Equation (\ref{eqn:Exp_model}).
The first two exponential terms represent the rising and decay of the fast component, respectively, and the last two account for the rising and decay of the slow component. \textit{A} and \textit{B} are the amplitudes of the fast and slow fluorescence components, respectively. $\tau_r$, $\tau_f$, and $\tau_s$ are the time constants of the rising edge, the fast light decay, and the slow light decay. $t_0$ is the time offset with respect to the acquisition window.

\begin{equation}
    L(t)=A\left(-e^{\left(-\frac{(t-t_0)}{\tau_r}\right)}+e^{\left(-\frac{(t-t_0)}{\tau_f}\right)}\right)+B\left(-e^{\left(-\frac{(t-t_0)}{\tau_r}\right)}+e^{\left(-\frac{(t-t_0)}{\tau_s}\right)}\right)
    \label{eqn:Exp_model}
\end{equation}

Gamma-ray-produced pulses exhibit mainly the fast light component \cite{BROOKS1979477}. Delayed fluorescence also exists, but its relative intensity is lower compared to prompt fluorescence. Therefore, the fit of the slow decay constant yields a large associated uncertainty \cite{ZHOU2022166287}. We hence set the amplitude of the slow component ($B$ in Equation \ref{eqn:Exp_model}) to zero and only fit the fast component of the gamma pulse template. 
After the fit, we calculated the difference between the measured template pulse and the exponential fit. The pulse time stamp of maximum difference can be considered as the time when the fit of the fast component fails to describe the complete gamma pulse because the model deliberately neglects the delayed fluorescence component. This timestamp represents the maximum of the delayed fluorescence component. Therefore, we chose it as the tail start time to start the tail integration of CI PSD.

\subsection{Figure of merit as PSD evaluation metrics }

\begin{figure}[htbp!]
    \centering
    \includegraphics[width=0.7\textwidth]{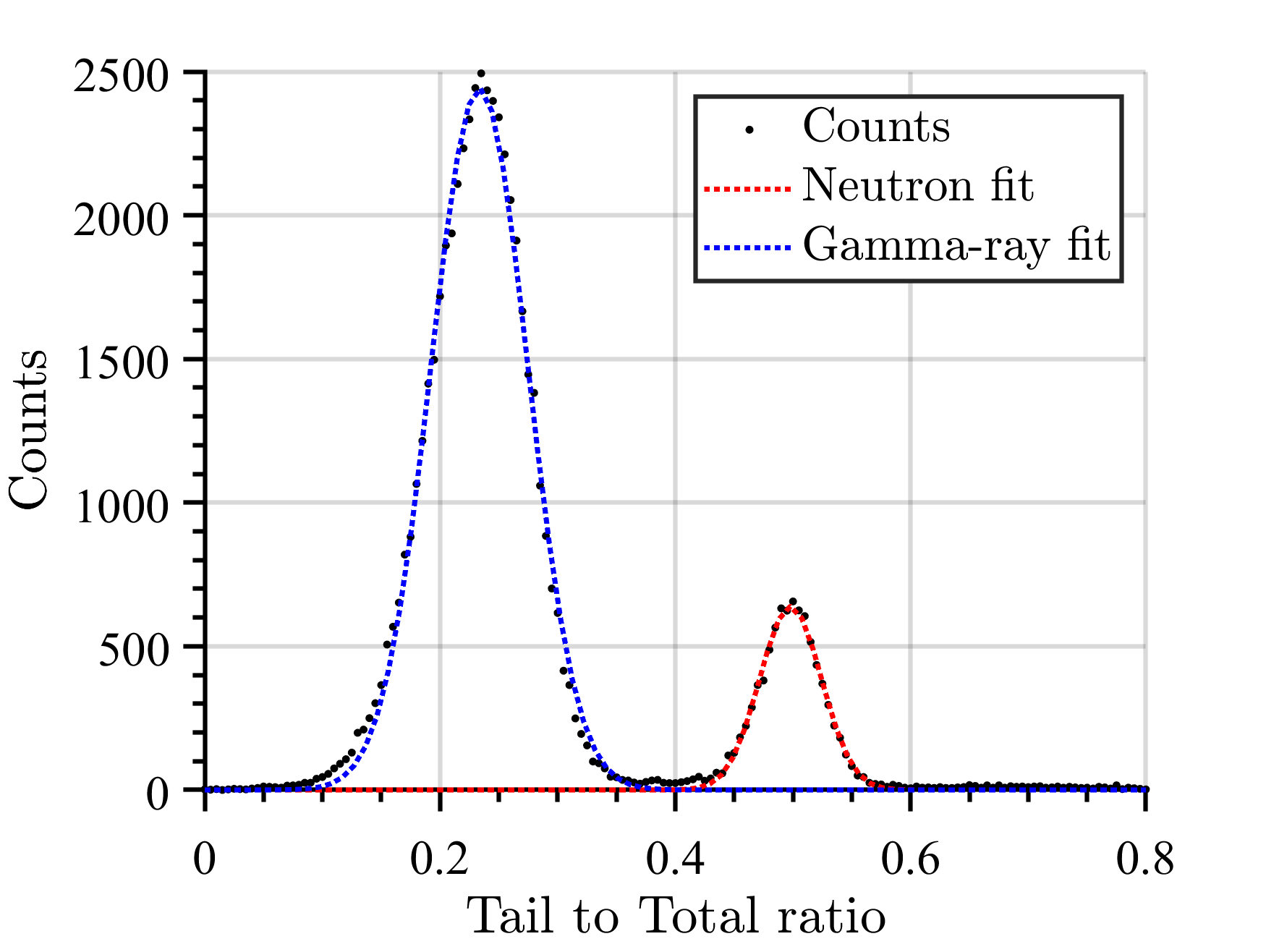}
    \caption{Distribution of $^{252}$Cf counts in terms of tail-to-total ratio in charge-integration-based PSD method, in the 200-210 keVee light output interval.}
    \label{f:m:fom_slice}
\end{figure}

\replaced{}{We used} The figure of merit (FOM), detailed below, \added{was used} as the quantitative metric to evaluate the PSD results. We performed charge integration PSD on each measured pulse and calculated the ratio of the tail and total pulse integrals (tail-to-total ratio). We analyzed the FOM of pulses in 20-keVee light output bins and evaluated the distribution of the trail-to-total ratio for each light output bin, as shown in Figure \ref{f:m:fom_slice}. The two peaks represent gamma and neutron pulse distributions, and their centroids and full width at half maxima (FWHMs) were used to calculate FOM, defined in Equation (\ref{eqn:FOM_eq}). \textit{S} is the distance between the maximum values of the neutron and the gamma-ray distributions, $\Gamma_1$ and $\Gamma_2$ are the FWHMs of the gamma-ray and neutron distributions, respectively. A larger FOM value represents a better neutron-gamma discrimination capability.

\begin{equation}
    FOM = \frac{S}{\Gamma_1 + \Gamma_2}
    \label{eqn:FOM_eq}
\end{equation}

\subsection{Experimental Setup}

\replaced{}{We demonstrated} The model-based PSD method \added{was demonstrated} on three organic scintillation detectors: a stilbene-d$_{12}$, an EJ-309, and an organic glass detector. Table \ref{t:scintillator_composition} shows the chemical composition and main properties of the detectors. The EJ-309 and the organic glass crystals are 5.08~cm tall cylinders with 5.08~cm diameter. The stilbene-d$_{12}$ has a 140~cm$^3$ non-equilateral hexagonal prismatic shape \cite{ZHOU2022166287} with a 5.4 cm height and was custom-grown at Lawrence Livermore National Laboratory \cite{Gaughan2021}. The three detectors are all coupled with photomultiplier tubes (PMT) that convert light pulses into electrical waveforms. The PMT models of stilbene-d$_{12}$, EJ-309 and organic glass detectors were HAMAMATSU H6559, ET Enterprises 9214B, and  ET Enterprises 9214B, respectively. These waveforms are acquired and digitized by a CAEN DT5730 500-MSps 14-bit digitizer. \par

\begin{table}[hbt!]
    \centering
    \caption{Material properties of the scintillators }
    \begin{tabular}{ |c|c| }
    \hline 
  Scintillator  & Composition \\
  & (wt $\%$)  \\
    \hline 
  Stilbene-d$_{12}$ &  46.15$\%$ deuterium, 53.85$\%$ carbon   \\
  EJ-309 & 55.52$\%$ hydrogen, 44.48$\%$ carbon    \\
  Organic glass & 45.59$\%$ hydrogen, 53.17$\%$ carbon, and 1.24$\%$ silicon \\
    \hline 
    \end{tabular}
    \label{t:scintillator_composition}
\end{table}

\replaced{}{We used} A 1$\mu$Ci $^{137}$Cs source \added{was used} to calibrate the detectors, in terms of light output. Then, we irradiated the detectors using a 5 $\mu$Ci $^{252}$Cf neutron source to evaluate their PSD performances. Each detector recorded approximately 2$\times$10$^{6}$ mixed neutron and gamma pulses emitted by the $^{252}$Cf source. The distance between the source and the front face of the detectors was set to 50~cm to have an acceptable detector count rate while minimizing the pile-up pulses. 
The data processing was performed with Python and Matlab custom codes.

\section{Results}
\replaced{}{We evaluated} The PSD performance of three organic scintillators \added{was evaluated} using the proposed PSD method. We calculated the FOM values obtained using the model-based method and compared it with the traditional PSD method based on the iterative optimization to find the integration time gates.
%\subsection{\textcolor{blue}{Model}-based PSD result of the stilbene-d$_{12}$ scintillator}

\subsection{Detector calibration and pulse template generation with the $^{137}$Cs source}

\begin{figure}[htbp!]
     \centering
     \includegraphics[width=0.7\textwidth]{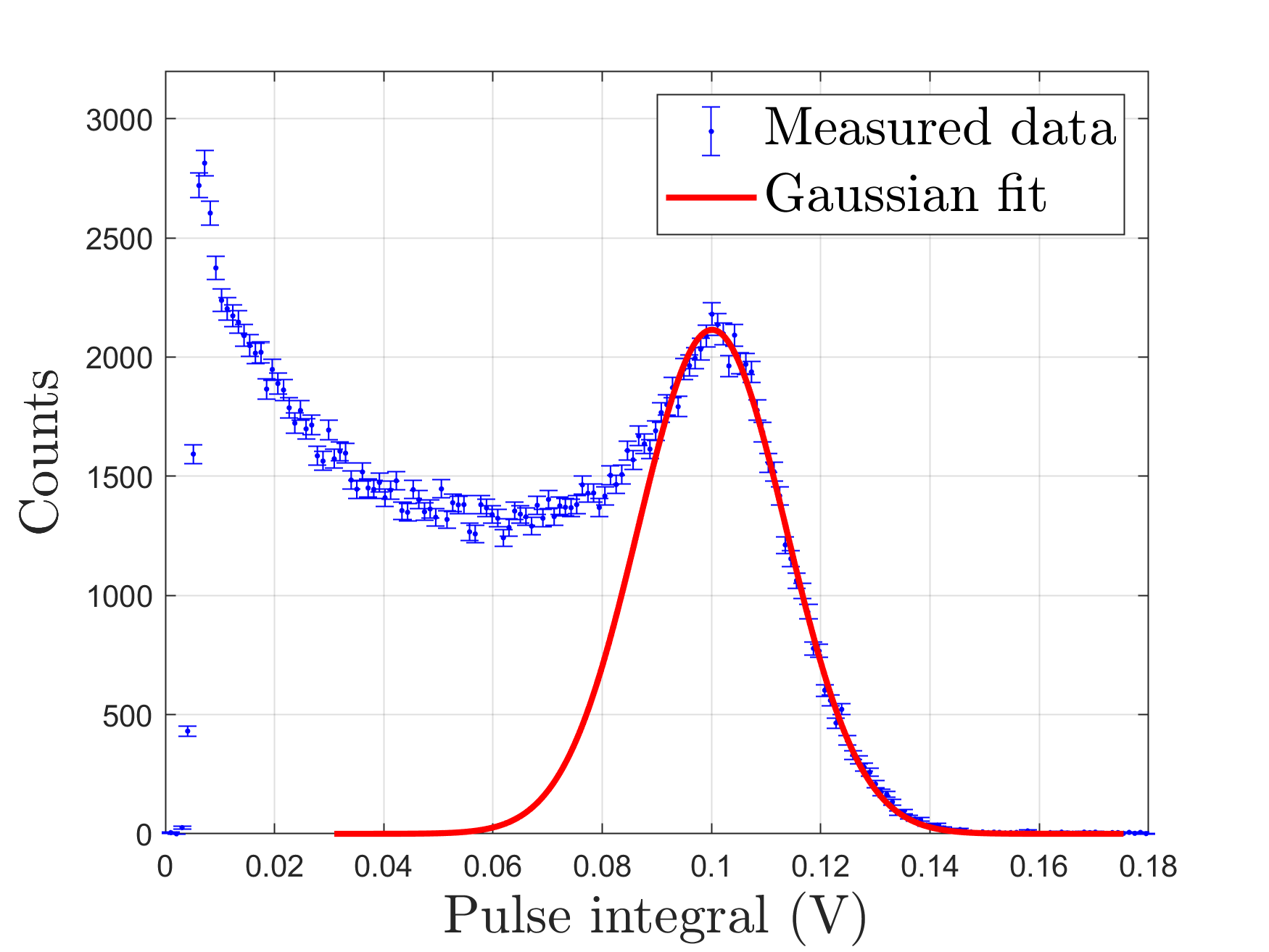}
     \caption{ Measured pulse height distribution of the stilbene-d$_{12}$ detector with the $^{137}$Cs source.}
    \label{f:r:dsb_cali}
 \end{figure}

We calibrated the stilbene-d$_{12}$, EJ-309, and organic glass detectors with the $^{137}$Cs gamma source. Figure \ref{f:r:dsb_cali} shows the measured $^{137}$Cs pulse integral spectrum with the stilbene-d$_{12}$. We fit the Compton edge with a Gaussian distribution and calibrated the light output response of the stilbene-d$_{12}$ to Compton electrons using the 85$\%$ of the edges produced by 662~keV gamma rays that correspond to 478~keV electron energy deposited via Compton scattering\cite{ZHOU2022166287}. The light output calibration of the stilbene-d$_{12}$ is: light output (keVee) = 4345.45 (keVee/V)$\times$Pulse height (V). We used the same calibration method for the EJ-309 and organic glass detectors, and the results are light output (keVee) = 2108.51 (keVee/V)$\times$Pulse height (V) and light output (keVee) = 1671.33 (keVee/V)$\times$Pulse height, respectively.

\begin{figure}[htbp!]
    \centering
    \includegraphics[width=0.7\textwidth]{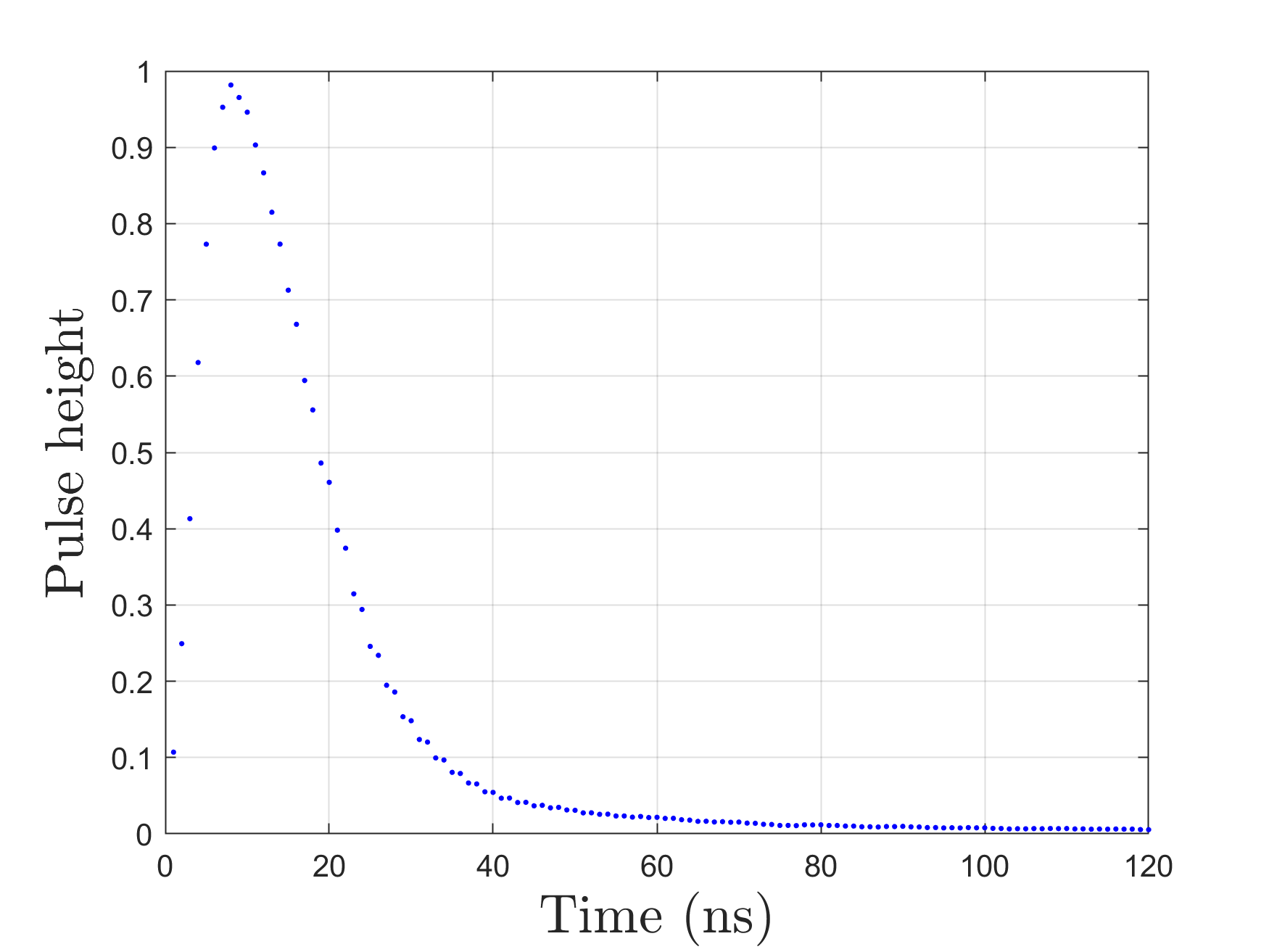}
    \caption{ Gamma pulse template obtained from the $^{137}$Cs measurement with the stilbene-d$_{12}$ detector.}
    \label{f:r:dsb_pulse_template}
\end{figure}

We also used the $^{137}$Cs measurement to generate the gamma pulse template. In order to ensure the quality of the pulse template, we rejected the piled-up pulses in the measurement. Approximately one thousand pulses whose pulse heights were within the $\pm$ 0.05V of $^{137}$Cs Compton edge region were chosen to build the template to reject low-amplitude pulses with high noise. The start of each gamma pulse was defined as the time when its amplitude reached 10$\%$ of its maximum. Then, 
we normalized the peaks of all gamma pulses to 1 and averaged them to form a pulse template, as shown in Figure \ref{f:r:dsb_pulse_template}. The time interval of two contiguous sampled points is 1~ns.

\subsection{Exponential fit of the pulse template and the acquisition of the tail start time from the fit result}

\begin{figure}[htpb!]
     \centering
     \begin{subfigure}[b]{0.5\textwidth}
         \centering
         \includegraphics[width=\textwidth]{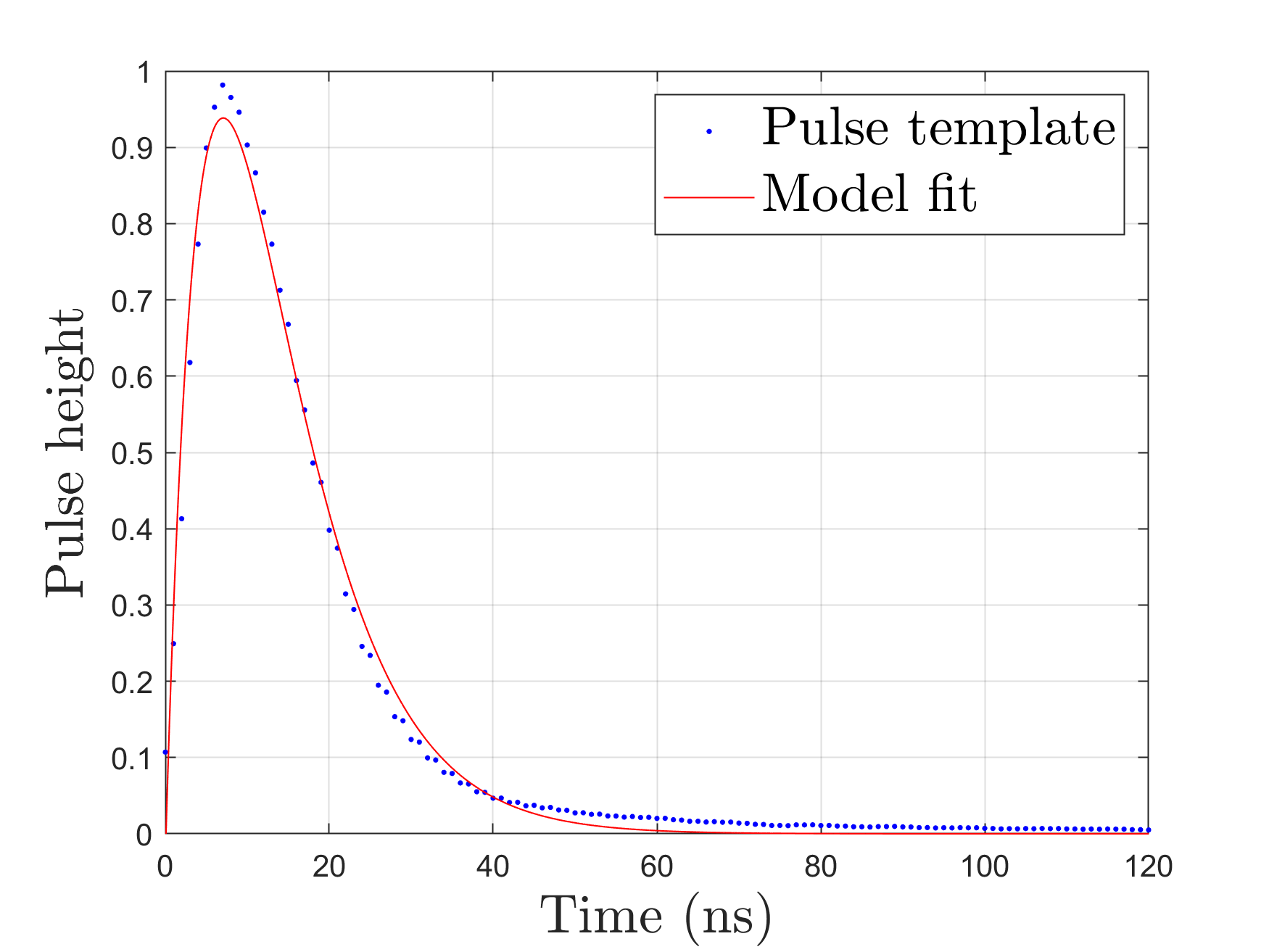}
         \caption{Exponential fit of the pulse template}
     \end{subfigure}%
     \begin{subfigure}[b]{0.5\textwidth}
         \centering
         \includegraphics[width=\textwidth]{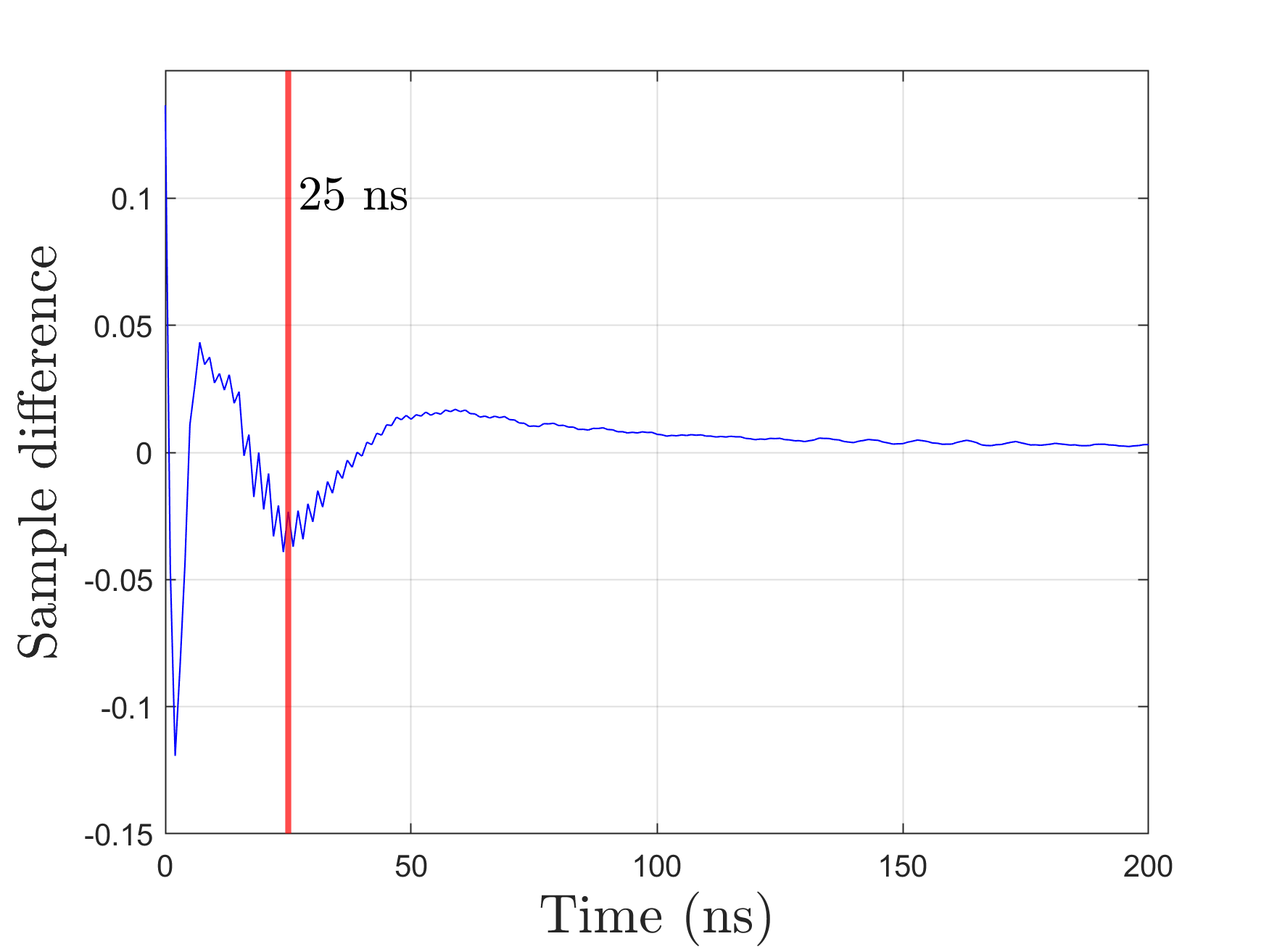}
         \caption{Difference between the template and the fit}
     \end{subfigure}
        \caption{Exponential fit result of the stilbene-d$_{12}$ gamma pulse template. }
        \label{f:r:dsb_pulse_fit}
\end{figure}

\replaced{}{We used}The exponential model (Equation \ref{eqn:Exp_model}) \added{was used} to fit the gamma pulse template. Figure \ref{f:r:dsb_pulse_fit} shows the fit result of the pulse template from the stilbene-d$_{12}$ detector. The fit was calculated using the curve\_fit function of the Python scipy.optimize package. 
Since we only used the exponential model to fit the fast component of the fluorescence signal, the fit does not resemble the template shape well at the tail region. This discrepancy allows identifying the onset of the delayed fluorescence component. Although the delayed fluorescence is relatively more intense in pulses produced by high-LET interactions, Compton electrons also exhibit a delayed fluorescence signal \cite{BROOKS1979477}. Figure \ref{f:r:dsb_pulse_fit} (b) shows the sample-by-sample difference between the pulse and the fit for the stilbene-d$_{12}$ detector. The maximum difference occurs 25~ns from the beginning of the pulse (18~ns from the pulse peak). We used this 18~ns from the peak as the tail start time to perform CI PSD for the stilbene-d$_{12}$. The tail start times of the EJ-309 and organic glass detectors were also obtained with this model-based method and they were 10~ns after the peak, as shown in Table \ref{t:CFD_sum}.

\begin{table}[hbt!]
    \centering
    \caption{Model-determined tail start time of three scintillators }
    \begin{tabular}{ |c|c|c|c| }
    \hline 
  Scintillator  & Stilbene-d$_{12}$ & EJ-309 & Organic glass\\
  & & & \\
    \hline 
  Fast decay time ($\tau_f$) from the fit & 7.9~ns& 4.3~ns & 3.9~ns \\
%  Relative light yield at 0.478 MeVee  & 1.0 & 1.32 & 1.65\\
  Model-determined tail start time  & 18~ns & 10~ns & 10~ns \\
%  FOM with the CFD method at 500-510 keVee  & 2.94$\pm$0.09 & 1.88$\pm$0.08 & 1.40$\pm$0.03 \\

%FOM with the proposed method & \multirow{2}{2em}{2.94}& \multirow{2}{2em}{1.88} & \multirow{2}{2em}{1.40}\\
% at 500-510 keVee  & & & \\
    \hline 
    \end{tabular}
    \label{t:CFD_sum}
\end{table}

%\begin{figure}[htbp!]
%    \centering
%    \includegraphics[width=0.7\textwidth]{figure/cfd_pulse.png}
%    \caption{ Gamma pulse template after the CFD process}
%    \label{f:r:dsb_cfd_pulse}
%\end{figure}

\subsection{PSD performance with the model-determined integration setting}

\begin{figure}[htpb!]
     \centering
     \begin{subfigure}[b]{0.5\textwidth}
         \centering
         \includegraphics[width=\textwidth]{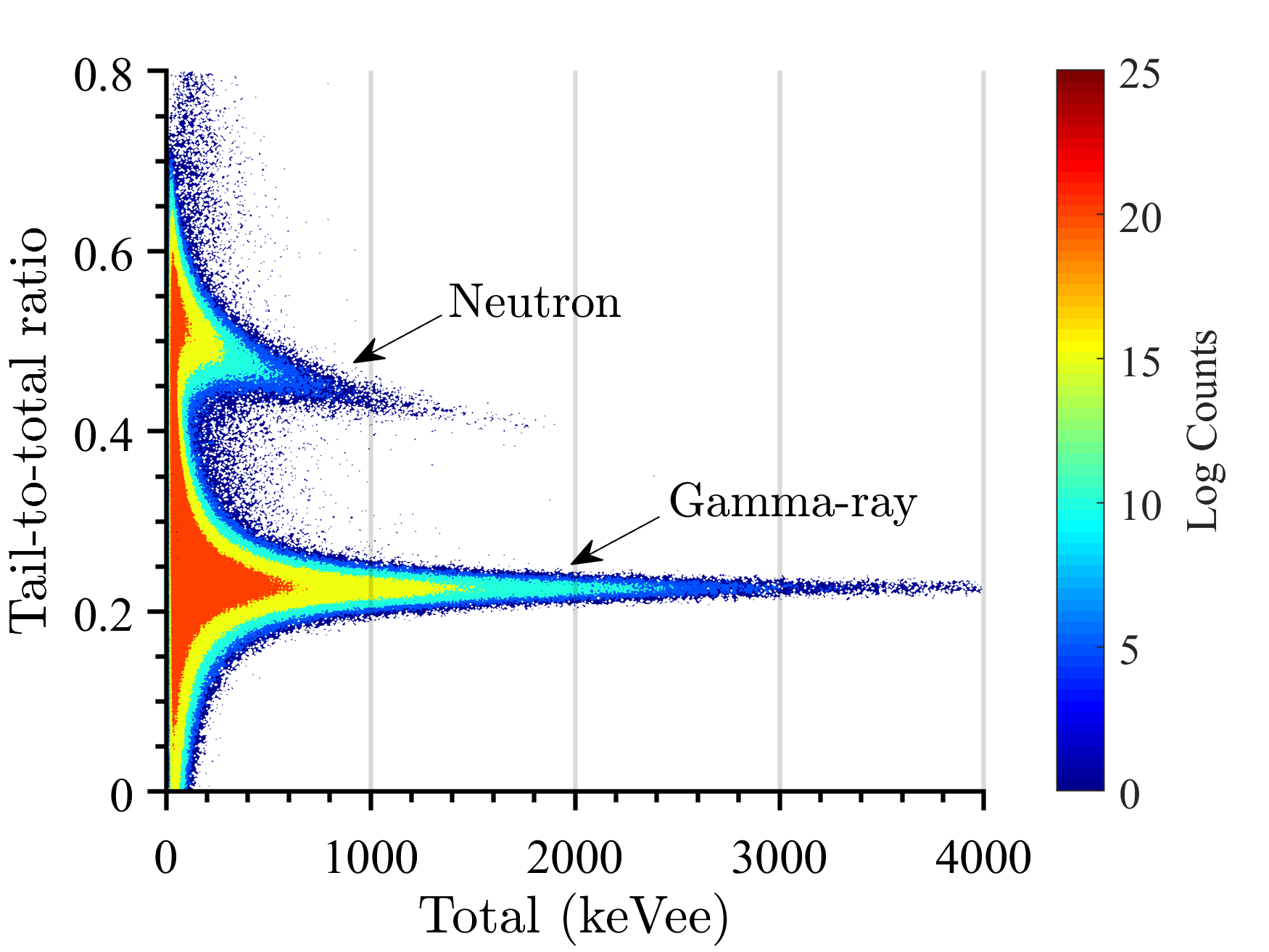}
         \caption{Stilbene-d$_{12}$}
     \end{subfigure}%
     \centering
     \begin{subfigure}[b]{0.5\textwidth}
         \centering
         \includegraphics[width=\textwidth]{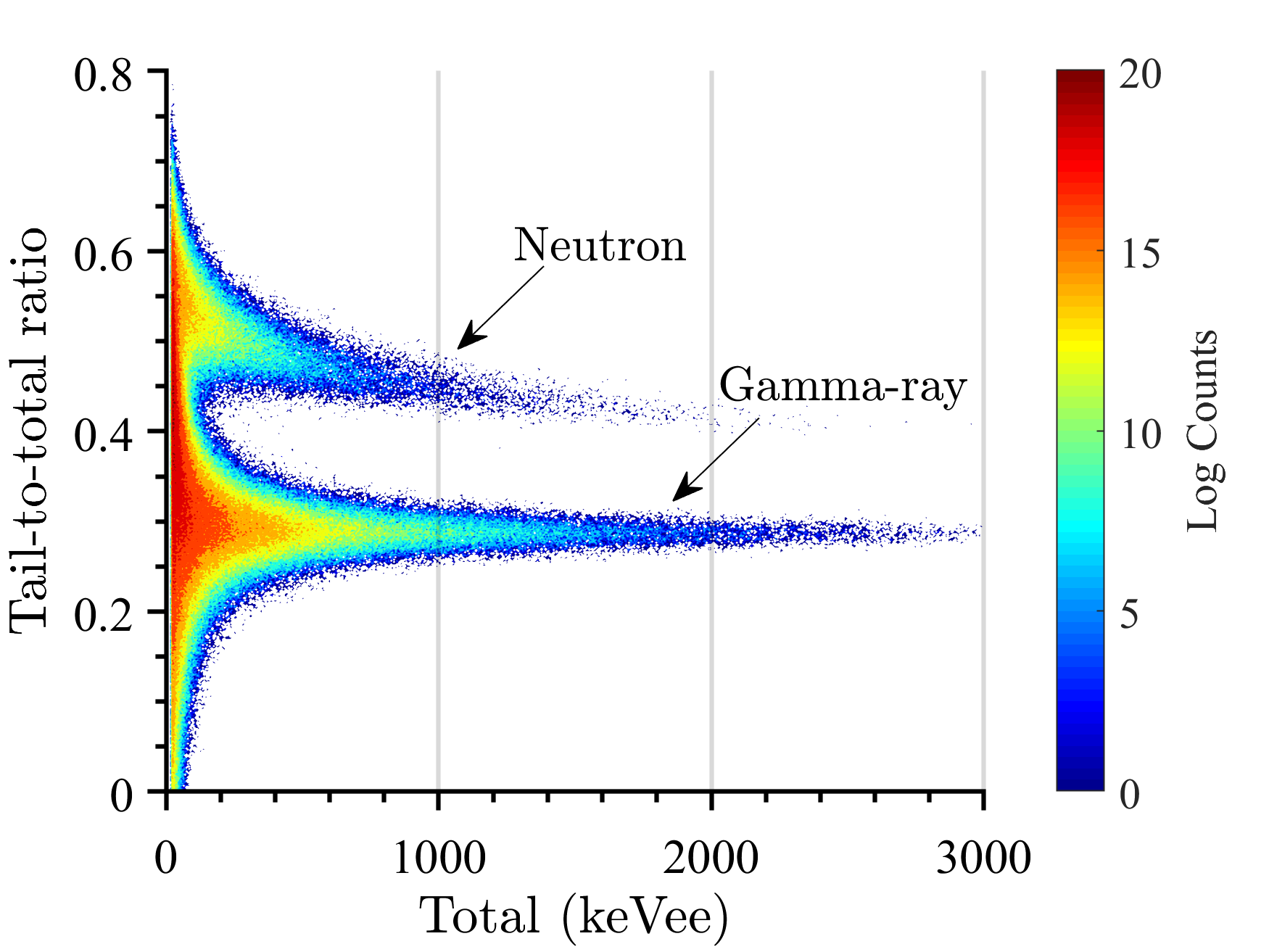}
         \caption{EJ-309}
     \end{subfigure}
     \begin{subfigure}[b]{0.5\textwidth}
         \centering
         \includegraphics[width=\textwidth]{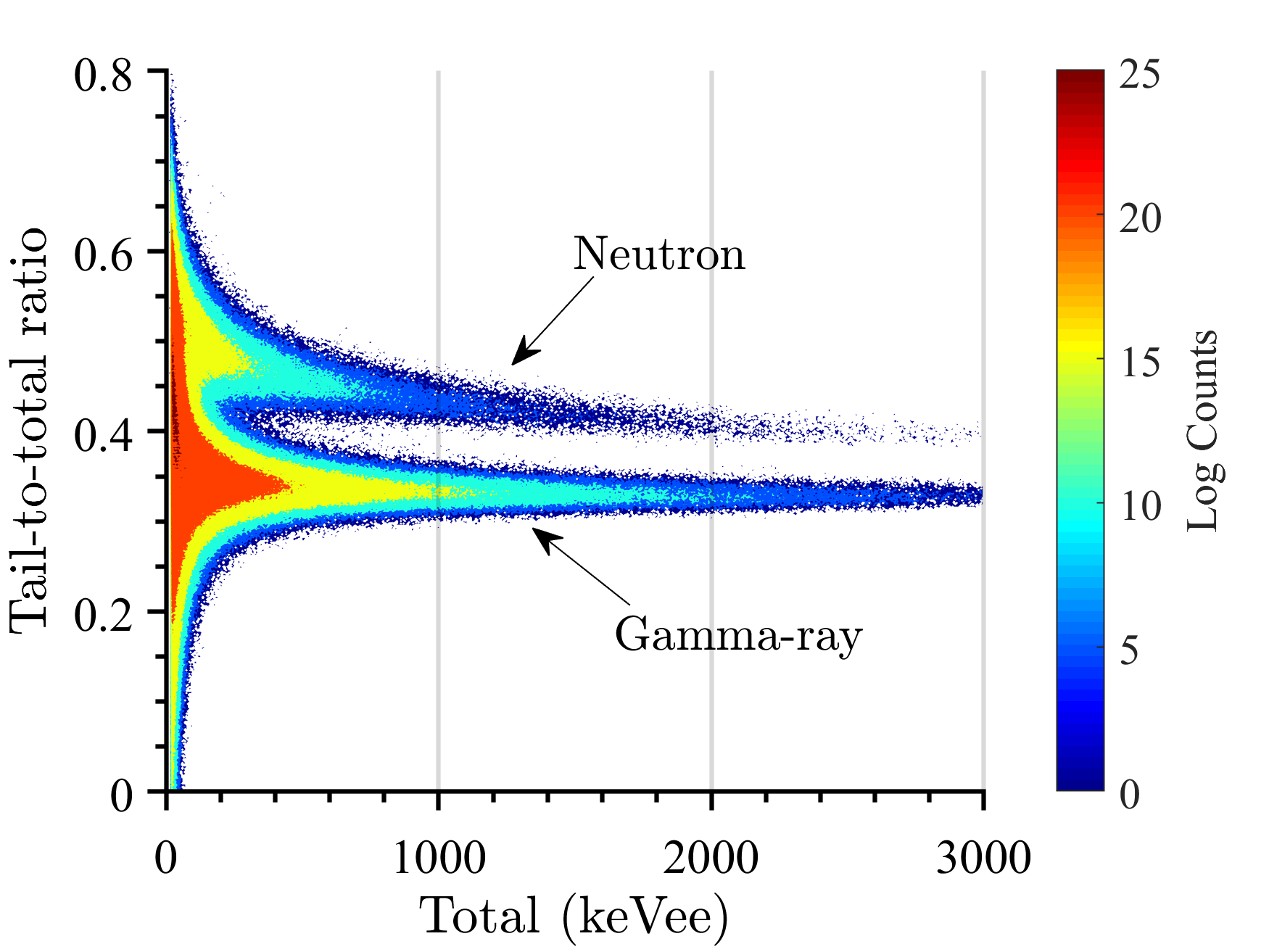}
         \caption{Organic glass}
     \end{subfigure}
        \caption{PSD scatter-density plot of the $^{252}$Cf source with model-determined integration setting. }
    \label{f:r:dsb_psd_plot}
\end{figure}

\replaced{}{We used}The model-determined tail start time \added{was used} to perform CI PSD for the stilbene-d$_{12}$, EJ-309, and organic glass detectors. The other charge integration settings were a) total integration started at 2~ns before the pulse peak, and b) the integration of the total and tail both ended at 150~ns. Figure \ref{f:r:dsb_psd_plot} shows the $^{252}$Cf PSD scatter-density plot of three detectors when using the model-determined tail start time settings. \replaced{}{We} One can observe that the neutron and gamma pulses are best separated in Figure \ref{f:r:dsb_psd_plot} (a), which demonstrated the stilbene-d$_{12}$ outperformed the other detectors in PSD. 
EJ-309 detector exhibited a better PSD capability than the organic glass detector. 

\subsection{ FOM sensitivity to the tail start time} 

\begin{figure}[htpb!]
    \centering
     \begin{subfigure}[b]{0.5\textwidth}
         \centering
         \includegraphics[width=\textwidth]{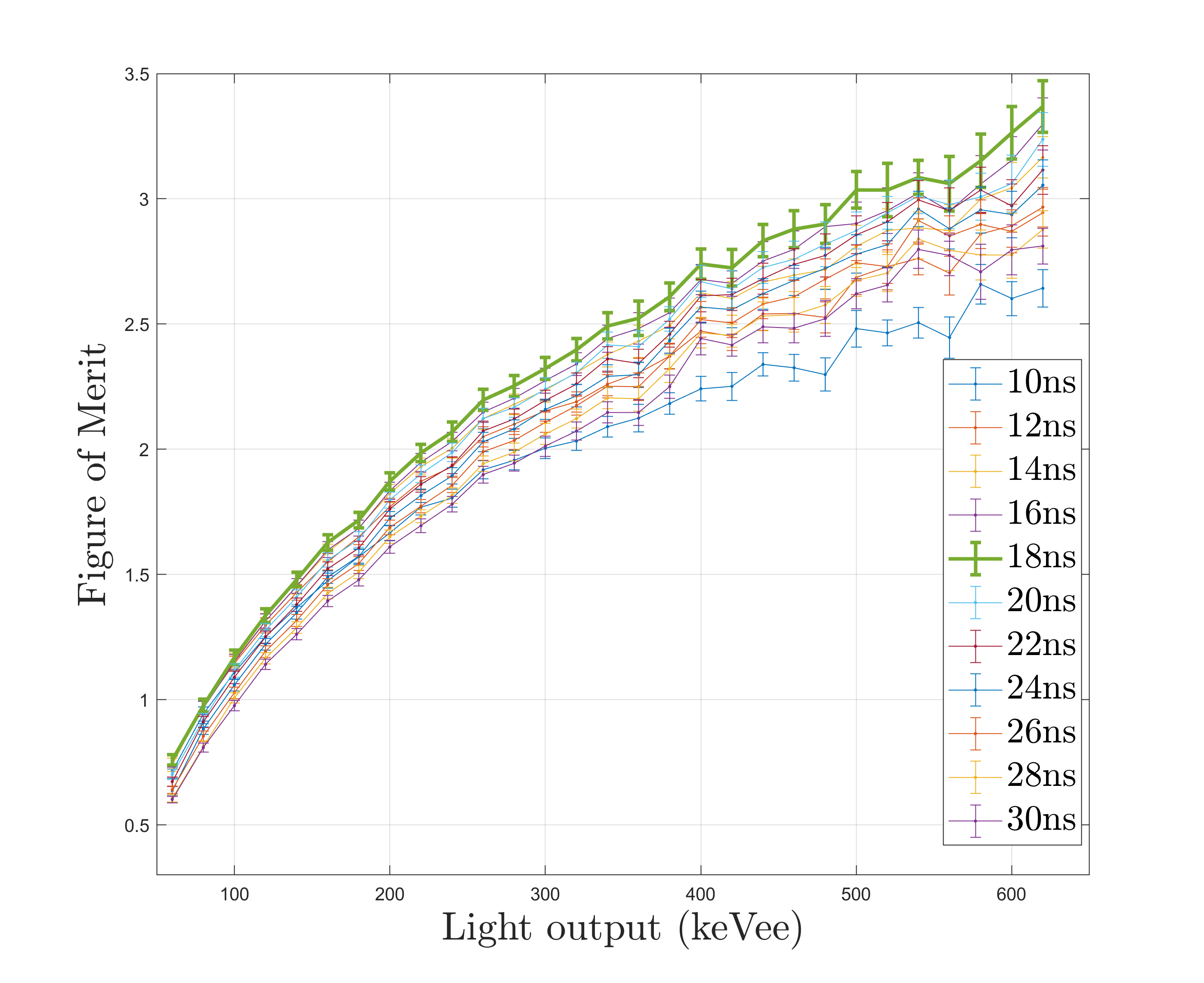}
         \caption{Stilbene-d$_{12}$}
     \end{subfigure}%
     \centering
     \begin{subfigure}[b]{0.5\textwidth}
         \centering
         \includegraphics[width=\textwidth]{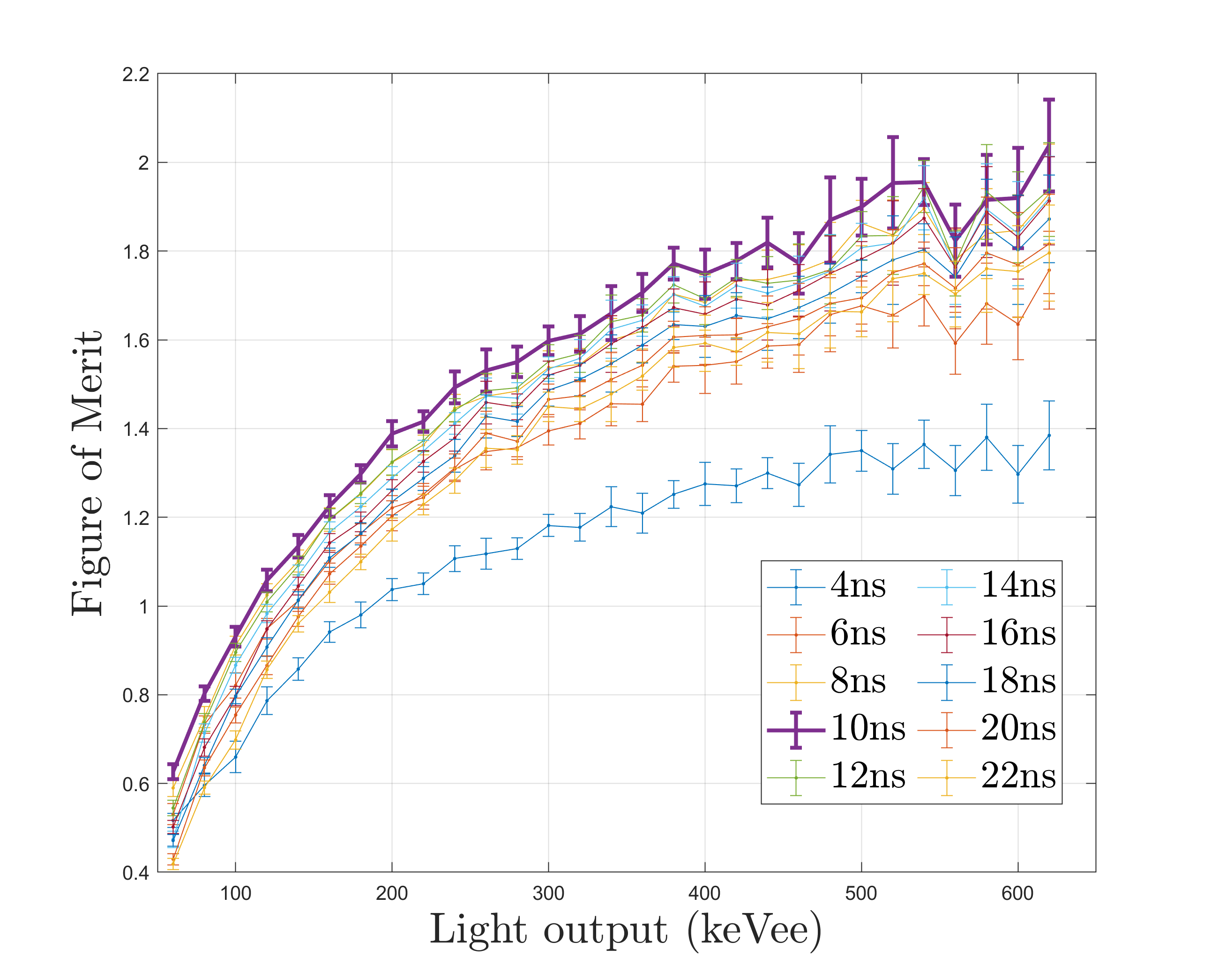}
         \caption{EJ-309}
     \end{subfigure}
     \begin{subfigure}[b]{0.5\textwidth}
         \centering
         \includegraphics[width=\textwidth]{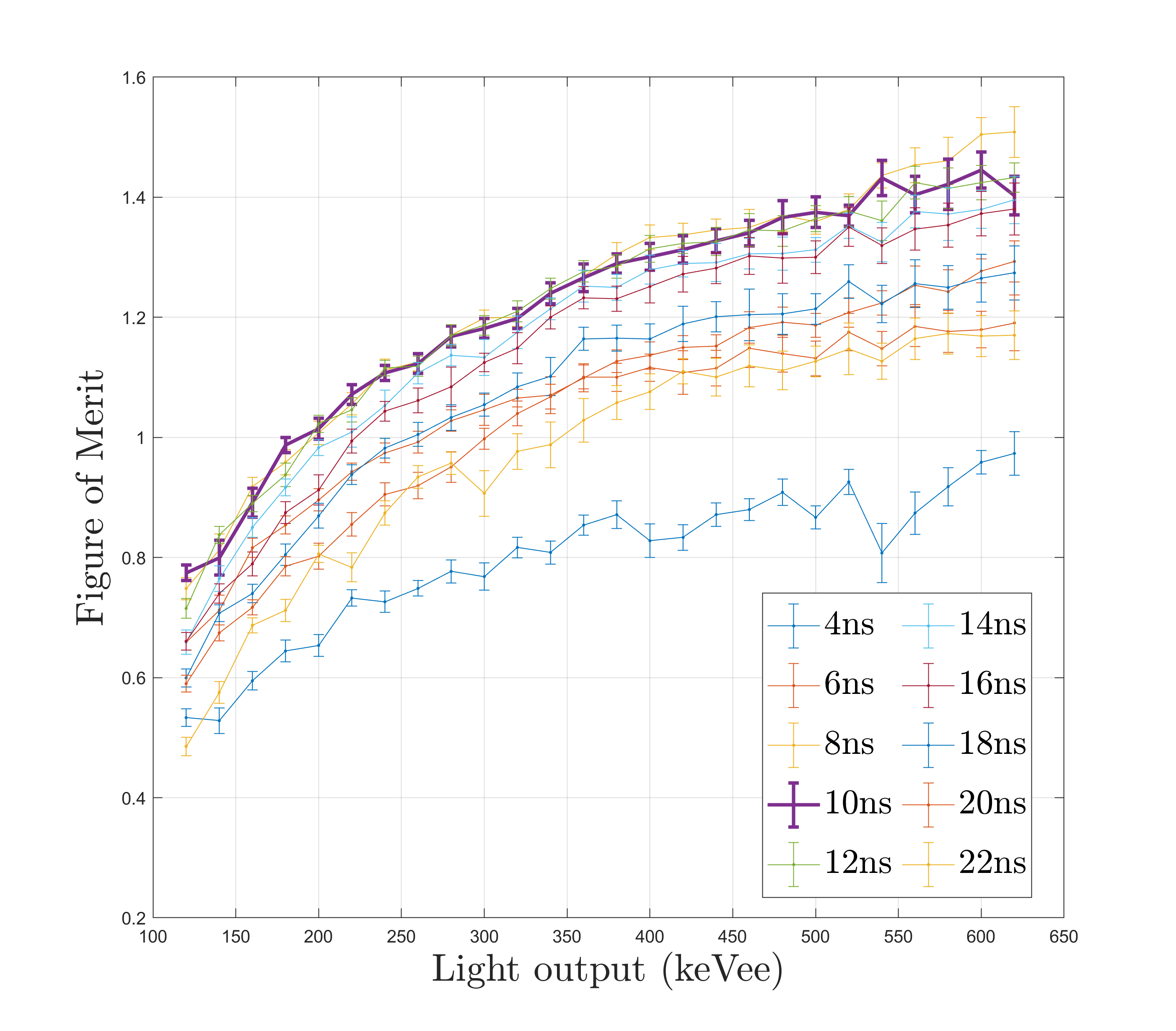}
         \caption{Organic glass}
     \end{subfigure}
        \caption{FOM of three detectors with various trail start time settings. The highlighted data represent the FOM values using the model-determined tail start time.}
    \label{f:r:fom_dsb}
\end{figure}

We calculated the FOM as a function of light output to quantitatively evaluate the PSD performance. Figure \ref{f:r:fom_dsb} shows the FOM of three detectors, when varying tail start time values. The tail start time values rang from 10~ns to 30~ns for the stilbene-d$_{12}$ and range from 4~ns to 22~ns for the EJ-309 and the organic glass. The other charge integration settings were kept the same (the pulse total integration started 2~ns before the pulse peak, and the total and tail integration ended at 150~ns). In Figure \ref{f:r:fom_dsb}, we can observe that the tail start time that was determined by the our method yields the highest FOM at all light output values.  \par
%The FOM values were calculated in the 50-600 keVee light output range.corresponding to a neutron deposited energy by deuteron recoils of 0.39-2.95 MeV in stilbene-d$_{12}$.}

In our previous work \cite{Gaughan2021}, we reported that the FOM value of a stilbene-d$_{12}$ crystal was approximately 3 at 400 keVee. In Figure \ref{f:r:fom_dsb}, the FOM value with the optimized integration settings is 2.74 $\pm$0.06 at 400 keVee. The stilbene-d $_{12}$ crystal in this paper is approximately 4.3 times larger than the one we used before \cite{Gaughan2021}. Light scattering effects within the crystal increase with its size and broaden the PSD distributions hence increasing the FWHM in \ref{eqn:FOM_eq} and worsening the PSD performance.\par

As for the FOM results of the glass detector, Laplace et al \cite{Laplace2020} reported the PSD performance of a 2 inches diameter and 2 inches height organic glass scintillator, and the FOM value is approximate 1.4 at 600~keVee. This result is in good agreement with the FOM value obtained in this work, 1.42 $\pm$0.03 at 600~keVee. Shin et al \cite{Shin201936} also reported the PSD performance of a 2 inches diameter and 2 inches height organic glass crystal, and found a slightly higher FOM of approximately 1.7 at 600 keVee. \par
As for the PSD performance of the EJ-309 detector, we obtained a 1.91$\pm$0.10 FOM values at 600~keVee. Using the same detector size and 600~keVee light output range, Laplace et al \cite{Laplace2020} reported an approximate 1.4 FOM value, Stevanato et al \cite{STEVANATO201296} reported an approximated 1.75 FOM value. 
Small discrepancies between our FOM and values reported in the literature could be due to different PMT models coupled to the scintillators, physical conditions of the crystals/glass and optical reflectors or impedance matching gel between the PMT and the crystal. Any optical phenomenon that affects the light propagation between the photon production at the position of interaction and its detection at the PMT photocathode could also lead to different PSD performance. Additionally, not all the cited papers reported the methods for PSD optimization, therefore, the cited PSD FOM parameters may have not been thoroughly optimized.

\subsection{FOM sensitivity to the total start time}
\replaced{}{We demonstrated that} The proposed model-based method can find the optimum PSD tail start time without the need of any neutron source or iterative algorithm. Besides the tail start time, the start time of the total integration is the other parameter that could affect the PSD performance. 
However, this parameter is expected to be consistent for the three measured scintillators because it does not depend on the fluorescence decay constants, which are material specific. \par
\begin{figure}[hp!]
     \centering
     \begin{subfigure}[b]{0.5\textwidth}
         \centering
         \includegraphics[width=\textwidth]{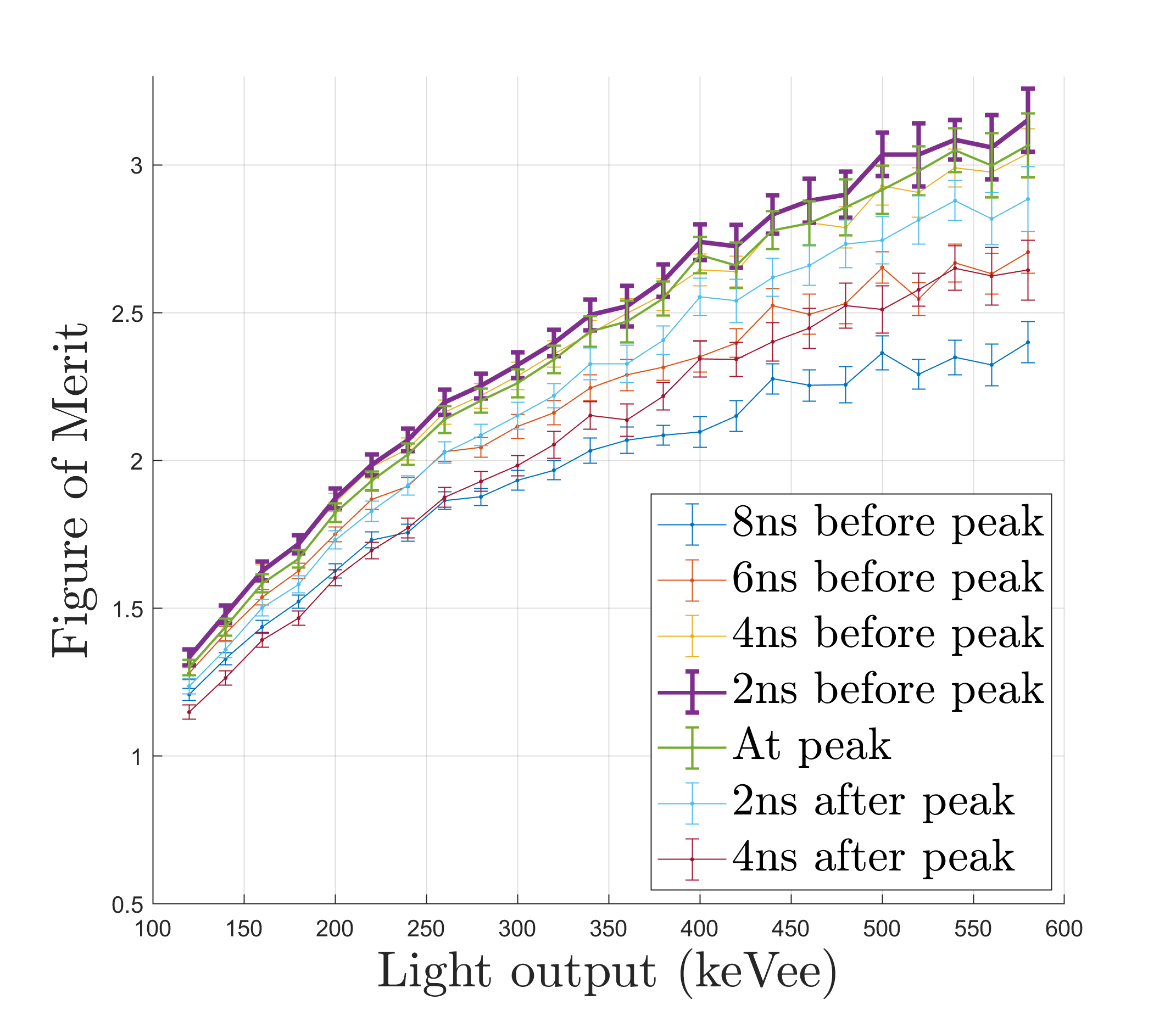}
         \caption{Stilbene-d$_{12}$}
     \end{subfigure}%
     \begin{subfigure}[b]{0.5\textwidth}
         \centering
         \includegraphics[width=\textwidth]{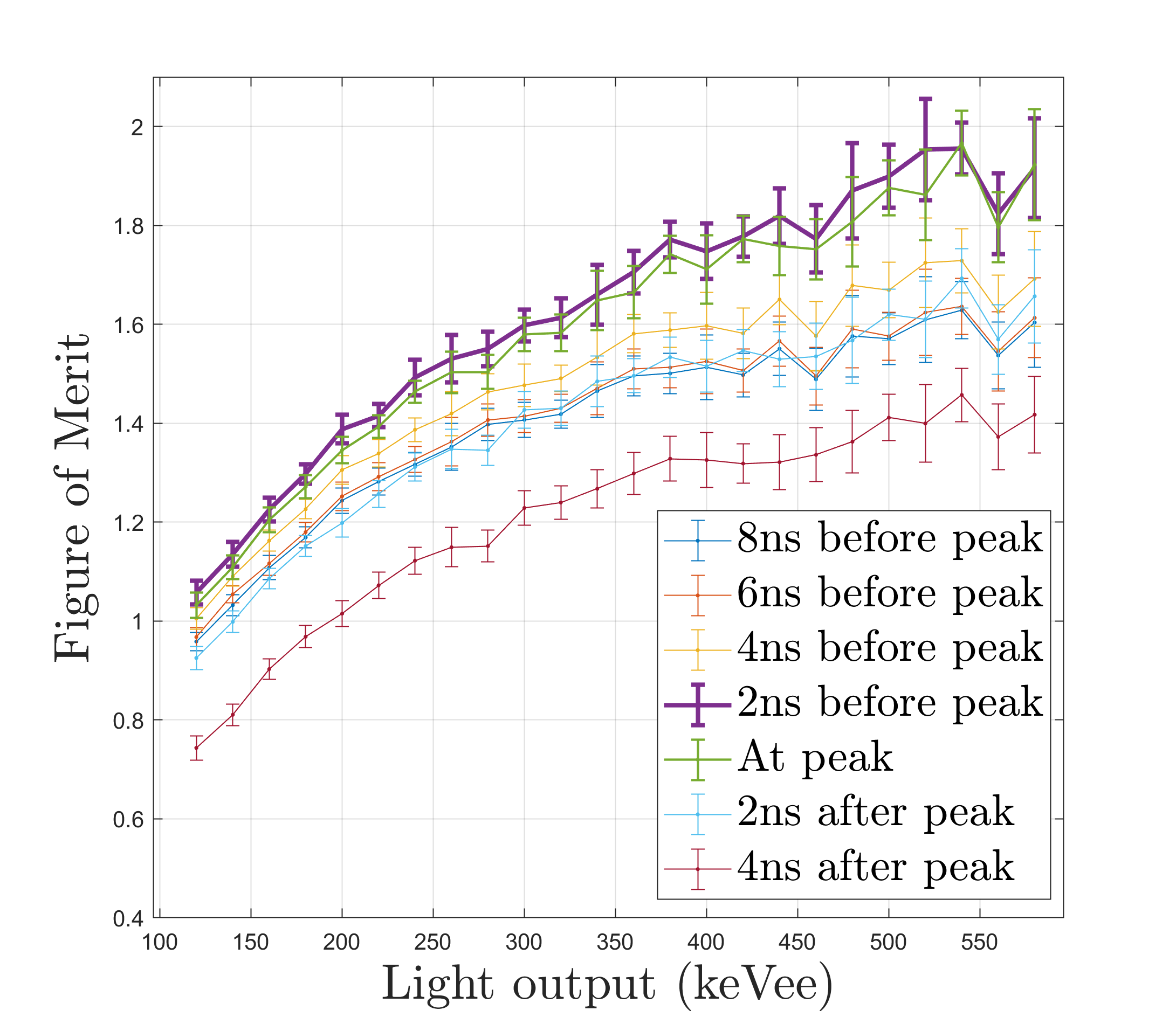}
         \caption{EJ-309}
     \end{subfigure}\hfill
          \begin{subfigure}[b]{0.5\textwidth}
         \centering
         \includegraphics[width=\textwidth]{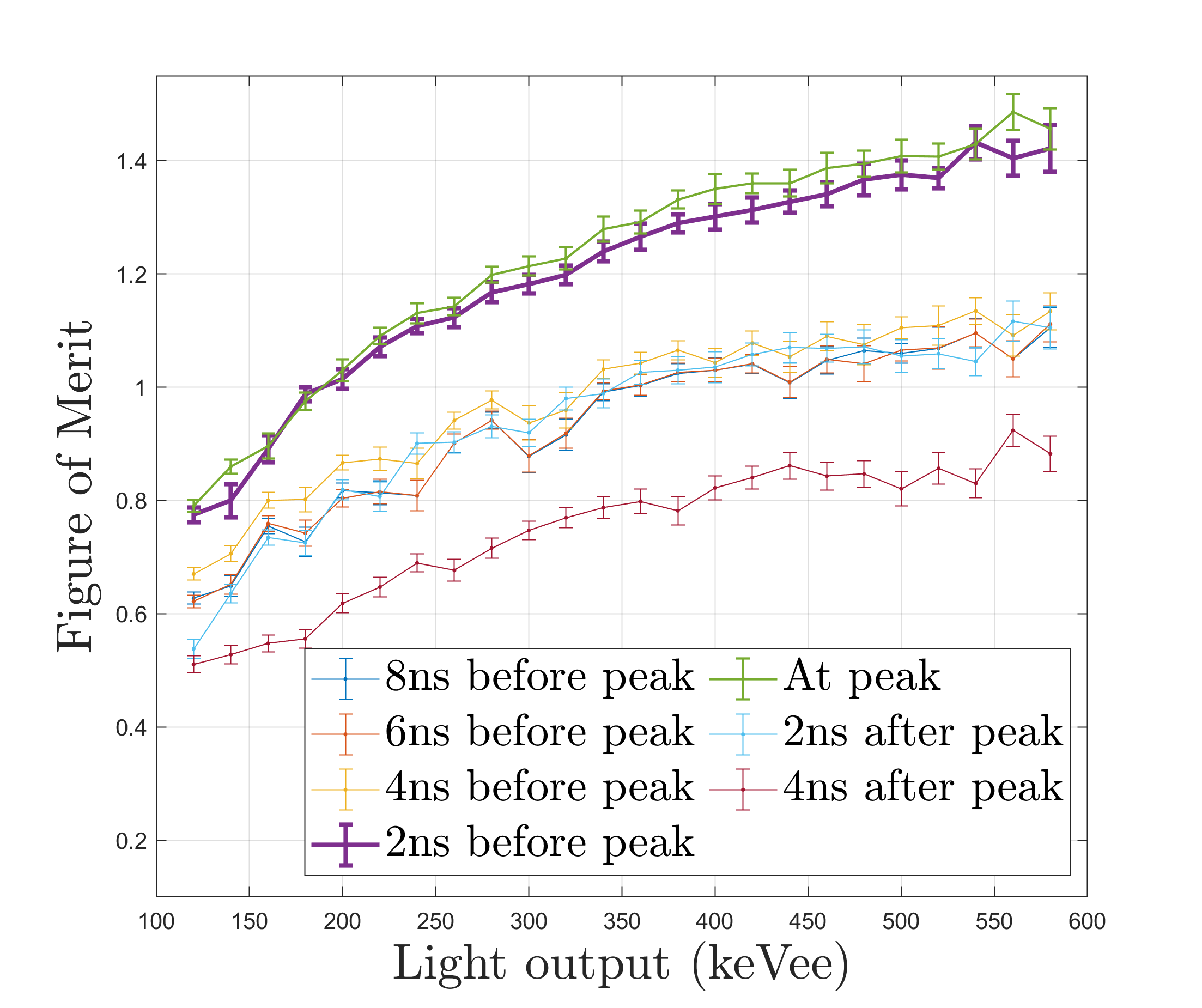}
         \caption{Organic glass}
     \end{subfigure}
        \caption{FOM of the stilbene-d$_{12}$, the EJ-309, and the organic glass detectors for the $^{252}$Cf measurement, using the model-determined tail start time settings. The legend represents the different total start times used in PSD.}
        \label{f:r:FOM_diff_total_start}
\end{figure}

We varied the total start time settings for all three organic scintillators and calculated the FOM values, while using the same model-determined tail start time setting. The results are shown in Figure \ref{f:r:FOM_diff_total_start}. We can observe that the PSD FOM is the highest when the total integration gate starts at 2~ns before the peak for the EJ-309 and stilbene-d$_{12}$, and slightly improves (within a single standard deviation) when the total gate starts at the peak for the glass detector. 

\section{Conclusions}
The optimization of PSD CI parameters is very important when high accuracy in radiation classification is needed, such as in nuclear reaction studies and nuclear non-proliferation measurements. 
CI time gate parameters need careful optimization to obtain the best discrimination between gamma-ray and neutron pulses. CI exploits the differential response in two different time gates of each detected pulse to derive a parameter that is shape-dependent and higher in pulses that exhibit a longer tail, i.e., more intense delayed fluorescence, with respect to faster-decaying pulses.
We showed that the parameter that influences the PSD FOM the most is the tail start time, as expected, being more sensitive to the fluorescence decay time constants. \\
We demonstrated that it is possible to optimize this parameter using a model-based method and a $^{137}$Cs source. The model-based method relies on the fact that the scintillation process in response to ionization interactions is always characterized by a prompt and one or multiple delayed fluorescence components. Even gamma-ray produced pulses, which are generally fast-decaying, exhibit all these components, but the prompt fluorescence is the prominent one. \\
We found the maximum discrepancy between a gamma-ray template pulse and its reduced exponential model, i.e., only including the fast component. 
The corresponding time stamp in the pulse indicates the onset of the delayed fluorescence component and can be used as tail start time. One could also fit the full model, including both fast and delayed components, but the goodness of the fit is poorer due to the relatively low intensity of the delayed fluorescence in gamma-ray pulses. %caused by triplet-triplet annhiliation whose probability is higher for high LET recoils, with a shorter range compared to low LET recoils (electrons produced by gamma rays of the same energy).
With this method, we found that starting the pulse tail at 18~ns, 10~ns, and 10~ns ns after the peak, in stilbene-d$_{12}$, EJ-309, and organic glass pulses, respectively, yields the best PSD FOM.
%These values are consistent with the slow pulse time constants of xx, xx, xx ns for stilbene-d$_{12}$, EJ-309, and organic glass, respectively.
The FOM obtained with the parameters that we determined are in good agreement with those found by other researchers and available in the literature, for organic glass and stilbene-d$_{12}$. We reported a higher EJ-309 FOM than Laplace and Stevanato \cite{Laplace2020, STEVANATO201296}. These discrepancies could be due to the digital pulse sampling time \cite{FLASKA2013456}, physical differences between detector and PMT setup, and to differences in the CI gate determination methods. This variability also confirms the PSD sensitivity to multiple physical parameters and the need of a robust PSD FOM maximization strategy. The method reported in this paper is simple to implement, does not require an iterative optimization process and can be performed with a common laboratory $^{137}$Cs source.

\section{Acknowledgements}

This work was funded in part by the Nuclear Regulatory Commission (NRC), United States Faculty Development Grant 31310019M0011 and in part by the Royal Academy of Engineering under the Research Fellowship scheme RF201617/16/31.

\medskip

\bibliography{reference}

\end{document}